# Using Web Page Titles to Rediscover Lost Web Pages

Jeffery L. Shipman, Martin Klein, and Michael L. Nelson

Old Dominion University, Department of Computer Science
Norfolk VA 23529

{jshipman, mklein, mln}@cs.odu.edu

**Abstract.** Titles are denoted by the TITLE element within a web page. We queried the title against the the Yahoo search engine to determine the page's status (found, not found). We conducted several tests based on elements of the title. These tests were used to discern whether we could predict a pages status based on the title. Our results increase our ability to determine bad titles but not our ability to determine good titles.

## 1 Introduction

There is a multitude of possibilities why a page or an entire web site may disappear [6]. These pages may reside in the caches of search engines, or web archives, or just moved from one URI to another [5]. There can only be one title in a web page. The title may not contain anchors, highlighting, or paragraph marks [8]. A 404 response code is an error message indicating that the client was able to communicate with the server but the server could not find what was requested.

Lost web pages and "404 Page Not Found" in many cases are not truly lost. A title of a web page may be used to recover pages returning the HTTP 404 response code. From a set of randomly selected URIs, we conducted a Yahoo search engine query based on the URI's title. Any page that did not return the associated URI within the top 10 results was considered lost. This paper discusses the probability of determining whether these page's status may be discerned from there titles or properties attributed to the title. The title or its properties are used to determine a good title from a bad one in using the title as a search string. The search for a page using the search string of "home page" returns over 9,000,000,000 results from a Yahoo search (Figure 1). If my goal was to find the URI `http://www.primaryecp.com/leveloptical/Home.aspx` which has the title of "Home Page" this type of search may not be productive. Using the same methodology, using the search string of "Tenet Group Home Page" to find the URI "`http://tenet.berkeley.edu/tenet.html`" yields the desire URI in the second position (Figure 2).

## 2 Experiment

### 2.1 Setup

At this time a collection of missing web pages does not exist. As such we need to create a set from the known web and pretend they are missing [7]. Given the fact that these URIs have be indexed by search engines, it can be concluded that querying by the right terms, will return them in a result set.

What would be most desirable for this experiment would be to take all URIs as our collection set. Regrettably, using the entire web as our test set is unrealistic. We were left with taking a random sampling. Capturing a representative sample set of web-sites for the entire web is not an insignificant task [11, 4]. Therefore, we selected a random collection of web pages from dmoz.org.

To choose a sample set, we selected a set of 7314 web-sites as our initial set of URIs. Our filters are similar to Park et al. [7]. We excluded URIs which contain non-English content from this sampling and all web sites with less than 50 words of textual content (HTML code excluded). This is necessary to allow for consistency within the set for future filters. Our final set consists of 7157 URIs. The distinction between found and not found was chosen as follows: A lost page was any page that was not within the top 10 results returned by the Yahoo search engine. The reasoning is most users actually will not look past a set amount of returned results.

### 2.2 Description

Taking a cursory look at the data set, one notices that when you examine web page titles by word count, the mean is 6.7, with a standard deviation of 3.3, giving a range of 3 to 10 terms (Figure 9). Similarly, examining web page titles by character count, the mean is the 44.7, with a standard deviation of 27.4, giving a range of 17 to 72 characters (Figure 8). Our data set was broken down into 66% found and 34% not found (Table 1).

Table 1. Characterize Search Returns

| Results   | Count | Percent |
|-----------|-------|---------|
| Found     | 4756  | 66.5%   |
| Not Found | 2401  | 33.5%   |

### 2.3 Method

The goal of the experiment is to discern an element or series of components within a title that would allow us to predict the status of a web page. If we summarily said all titles are good titles for our response, we would be correct 66% of the time. Our baseline or point of reference for determining if a test merits consideration, is a test that can discern good titles from bad titles more than 66% of the time. For a more precise examination of significance, the Fishers exact test will be used to compare a test's response to the baseline response [1, 2]. Fishers exact test is referred to as a statistical significance test. This is a test to determine that a relationship is unlikely to have occurred by chance. The test is meant for data that may be categorized in two different ways. The p-value from a Fisher test indicates the significance of a result. A low p-value means a low likelihood that the result occurred by chance hence the null hypothesis can be rejected hence the result is considered significant. The lower the p value, the more confident you can be that your result is significant. P of less than 0.05 is a common threshold to determine significance. The Fisher Exact test were performed using R, a program for statistical computation and graphics [3]. It consists of a language, plus a run-time environment with graphics, a debugger, access to certain system functions, and the ability to run programs stored in script files.

Our experiment focused on several aspects of a title. Given a title is a sentence like structure, we focused on the different aspects of the sentence. These include nouns, verbs, articles, adverbs, prepositions, and adjectives.

The next avenue of interest was stop words. Stop words is a term coined for words that do not add extra meaning to a search or process [10]. For this reason, search engines in general will dismiss search terms that are in the respective stop word set. The thought was that the more stop words present within a title, the less meaningful or helpful the title would be. Thus we hypothesized that the more stop words in a title the less likely the correct URI would be returned with a title submission, respectively.

The next test used was based on the search. A search for a particular title is a composition of two parts: the title and the type of search. Searches may be quantified into boolean OR search, boolean AND search, or quoted. Depending upon the boolean search used, effects the amount of results returned. Using these three types of searches as building blocks to determine singularly or in combination if these searches would lead to the discovery of good titles.

A boolean OR search is one in which the user enters the title as is,

<p align="center">Jeffery Shipman's home page</p>

This the equivalent to a boolean OR search. Results are returned that have one or more of the terms in the respective string. Similarly an boolean AND search is one in which each word is prefixed with a +,

<p align="center">+Jeffery +Shipman's +home +page</p>

This is the equivalent of a boolean AND search. Results which are returned have all of the terms in the respective string. The final search type is a quoted search.

<p align="center">"Jeffery Shipman's home page"</p>

Results which are returned have all the terms in the respective string and in the order they were presented to the search engine.

We used these boolean searches as building blocks for cluster analysis or clustering. Clustering or cluster analysis is the assignment of a set of observations into subsets (clusters) [9]. These clusters are similar to one another through one or more elements. We graphed the results returned based on the type of search performed and separating the queries by titles that the Yahoo search engine was able to find from those it was not.

The Final approach was considered after alphabetizing the set of titles. There was an obvious repeating of several titles. These selected titles produced poor results from the Yahoo search engine. With this knowledge, these "stop titles", were searched for as a Percentage of the title. Given that a title can be thought of as a grouping of words or an array of characters, a two prong approach was used. The first prong was consideration by words. A word in this experiment was deemed to be any series of characters separated by white space. The second prong was to consider the title as a series of characters (Table 2).

**Table 2.** Example Titles and Their Respective Counts

| Title | Word Count | Character Count |
|---|---|---|
| funky country.com | 2 | 17 |
| index of /bandbeastrunton@btinternet.com | 3 | 40 |
| welcome to my home page | 5 | 23 |
| welcome to my home page. | 5 | 24 |
| welcome–to–m::home*page | 1 | 26 |
| hi welcome to my home page | 6 | 26 |

## 3 Results

After reviewing the results for tests based on grammar related tokens, it is clear that the choice of titles with less than 13% adverbs produce the best outcome. The result for this type of search returned an accuracy of 66%. This means that the test was able to determine a good title, one that will return the URI, from a bad title, 66% of the time (Table 3). The difficulty with this result is that adverbs are represented poorly within the data set. After performing the Fisher exact test in R, we receive a p-value of 0.9718. This informs us that we have a result that can not be definitively proven to be significant when comparing against the baseline of accepting all titles as good titles. A more thorough examination of the parts of speech tests are listed in the appendix.

For the Stop Word based tests, the best outcome was 60%. This was produced by collecting a large corpus of accepted stop words and filtering out titles with a content of less than 35% stop words (Table 4).

**Table 3.** Set of adverb stop words divided by number of words less than 0.13

|  | Actual | | Total Mismatch | Percent Mismatch |
|---|---|---|---|---|
|  | found | not found | 2398 | 34 |
| **Predicted** found | 4746 | 10 |  |  |
| not found | 2388 | 13 | Match | Match |
|  |  |  | 4759 | 66% |

**Table 4.** Super set of traditional stop words divided by number of words in URI's title less than 0.35

|  | Actual | | Total Mismatch | Percent Mismatch |
|---|---|---|---|---|
|  | found | not found | 2856 | 40 |
| **Predicted** found | 3574 | 1182 |  |  |
| not found | 1674 | 727 | Match | Match |
|  |  |  | 4301 | 60% |

After performing the Fisher exact test in R, we receive a p-value of 3.395e-15. This informs us that we have a significantly worse result than just saying all titles are good titles. This may be due to the lack of accepted stop words in the set. The most prevalent words in the set were: and; the; of; home; to; welcome. The ubiquitousness of the words may have detracted from there usefulness. Additionally, words not considered stop words appeared in the top ten of most represented words such as home and welcome (Figure 3). In all of the designed tests, the element being searched for did not exist in enough significance to give confidence in the validity of a single process (Figure 4).

For query based tests, we conducted an array of different queries and combination of queries in hopes of producing clustering (Table 5). We graphed these results based on found and not found versus amount of results returned. No clustering was evident (Figure 5). An extensive list of graphs for query based tests may be found in the appendix.

The most significant finding was that there exists a set of titles that may be summarily removed from any search set. These titles are "stop titles". They may be titles formed due to the advent of "cookie cutter" web-sites, cloning of web-pages, web-site creating applications or web-site creating services (Figure 6). Analyzing titles based on the percentage of the title by word that was a stop title produced a 72% success rate (Table 6). Analyzing titles based on the percentage of the title by character that was a stop title produced a 72% success rate (Table 9).

**Table 5.** Search Results Tests

| Yahoo search of title quoted |
| --- |
| Yahoo search of title "and" |
| Yahoo search of title "or" |
| Yahoo search of title "or" versus Yahoo search of title "Quoted" |
| Yahoo search of title "and" versus Yahoo search of title "Quoted" |
| Yahoo search of title "or" divided Yahoo search of title "Quoted" |
| Yahoo search of title "and" divided Yahoo search of title "Quoted" |

**Table 6.** Found stop title divided by number of words in URI's title greater than 0.7

|  | **Actual** |  | Total Mismatch | Percent Mismatch |
| --- | --- | --- | --- | --- |
|  | found | not found | 2039 | 28 |
| **Predicted** found | 4753 | 3 |  |  |
| not found | 2036 | 365 | Match | Match |
|  |  |  | 5118 | 72% |

**Table 7.** Most Prevalent Stop Titles in Our Set

home
index
homepage
hometown has been shutdown - people connection blog: aim community network

The usability of a title becomes less as the title becomes more similar to the "stop titles" (Table 7). The difficulty becomes that one word may take a title from being of no value to one in which the URI may be returned in the search results. As an example of this problem is the title "my home page". It is 2/3 stop title (home page) by word. The same may be stated for "linuxguru home page" but the proper URI will be returned in the search results (Table 8).

Once titles that are "stop titles" are removed from the set of URIs, there is a visible transition between titles of distinct lengths. "Stop titles" are 5% of the total set. Titles that are ten words or less return the proper URIs in the top 10 listings 71% of the time. This subset contains 78% of the total data set. Titles that are between eleven words and twenty words return the proper URIs 65% of the time. This subset contains 14% of the total data set. Titles that are between twenty-one words and fifty words have a 44% chance of finding the proper URI. This subset contains 1% of the total data set.

**Table 8.** Characterization of URI Titles by Words, Characters, and Stop Title

| Title | Word Count | Character Count | % Stop Title Word | % Stop Title Char |
|---|---|---|---|---|
| my home page | 3 | 12 | 66% | 75% |
| linuxguru home page | 3 | 19 | 66% | 47% |

**Table 9.** Found stop title divided by number of characters in URI's title greater than 0.55

|  | **Actual** |  | Total Mismatch | Percent Mismatch |
|---|---|---|---|---|
|  | found | not found | 2038 | 28 |
| **Predicted** found | 4748 | 8 |  |  |
| not found | 2030 | 371 | Match | Match |
|  |  |  | 5119 | 72% |

After fifty words the likelihood of finding the desired URI drops to 18% (Figure 7). This subset contains less 1% of the total data set. This characterization of the data is shown in Table 12. Thus increasing the amount of search terms in a title does not empirically increase the chances of finding the searched for URI. This may be due to the lack of terms specific enough to return the desired URI while similarly produces an array of non-sensical URIs. This lack of specificity can be due to repeating words, the use of words that are too general in meaning, or the use of words that are used to often within the corpus of the web.

Our title of 101 words had several words repeating within the title

```
focustribe studios --- building brand innovation --- 949 258
0118 --- creative branding, web development, online marketing
--- web design, web applications, web strategy, user interface,
flash application, content management, enterprise ecommerce,
portal application, intranet portal, extranet portal, database
design, database development, business intelligence, e-learning,
product simulation, configurator, web application, ci, corporate
identity, logo design, corporate collateral, graphic design,
event marketing, tradeshow marketing and design, direct mail
campaigns, promotional cd-roms, copywriting services, email
marketing, search engine optimization, banner development,
advertising, online advertising, pay-per-click consulting,
focustribe studios, focustribe, focus121, focusone2one,
focusbrand, focussolutions, martina juchli, roland
schertenleib, juchli, schertenleib, newport beach, aliso
viejo,
```

The above title contains 26 duplicate words. The word "web" having five instances. The word "design" having five instances. The word "marketing" having four instances. A more detailed breakdown may be found in Table 10.

**Table 10.** Count of Duplicate Words for Title of 101 Words

| Count | Word |
|---|---|
| 5 | web |
| 5 | design |
| 4 | marketing |
| 3 | portal |
| 3 | focustribe |
| 3 | development |
| 3 | application |
| 2 | studios |
| 2 | schertenleib |
| 2 | online |
| 2 | juchli |
| 2 | database |
| 2 | corporate |
| 2 | advertising |

## 4 Future Work

Future work that may be considered includes a larger data set. This would most likely increase the discovery of more "stop titles" and possible a token that is more representative of a "good" title. Secondly, Apply this process to studies in non-English titles. Finally, Apply this process to a more extensive examination of stop words with respect to found and not found searches.

## 5 Conclusions

We randomly selected URIs from `dmoz.org` to create our base set. From this set we exclude all non-English URIs and all web-sites with less than 50 words of textual content We conducted a Yahoo search engine query based on the URI's title. We performed an array of tests on the URI's to discern whether we could produce better than 66% accuracy. This includes: tests based on grammar related tokens of a title; tests based on different types of searches; tests based on excluding different sets of stop words; tests based on excluding URI's with respect to the percentage a URI title was a stop title.

Our analysis of the data has shown that the usefulness of a titles for discovering a good titles is limited. This is most likely due to the difficulty in discovering words that are significant. The discovering of significant words is complicated by finding words that are unique enough to be useful but not so used in the web to have a large recall. Given this, we have shown that by excluding stop titles we increase the accuracy of discerning a good title from a bad title. This gain is increased with the realization that titles with larger amounts of words fair far worse than titles of ten words or less (Figure 7).

# A  Appendix

## A.1  Tests

Table 11. Test for Processing Titles

| Tests |
|---|
| length of title based on characters in a title |
| length of title based on words in a title |
| longest word within a title |
| yahoo search of title quoted |
| yahoo search of title "and" |
| yahoo search of title "or" |
| number of nouns in a title |
| number of adverb in a title |
| number of adjectives in a title |
| number of preposition in a title |
| number of articles in a title |
| number of stop words in a title |
| percentage stop title in a title with respect to words |
| percentage stop title in a title with respect to characters |

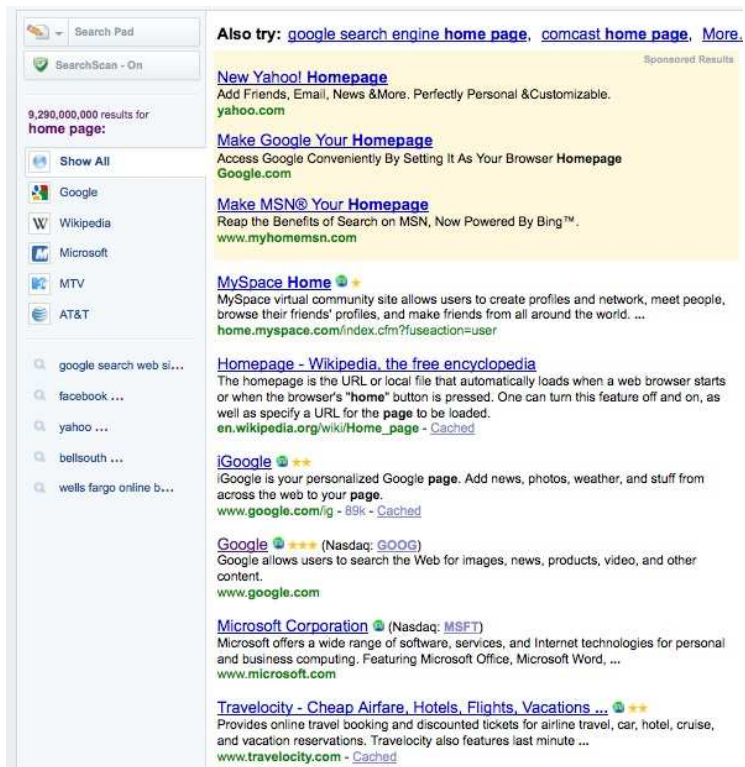

**Fig. 1.** Snap Shot of Yahoo Search using "Home Page" as search string

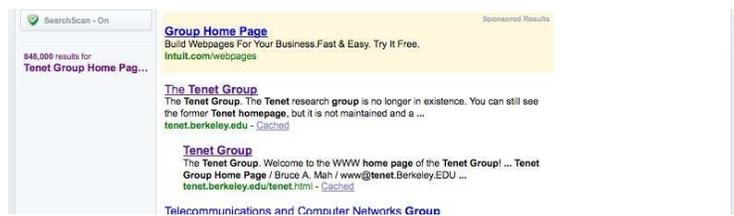

**Fig. 2.** Snap Shot of Yahoo Search using Tenet Group "Home Page" as search string

*Top 50 Term Ocurrances in Title's Dataset*

**Fig. 3.** List of Top Frequent Terms within set

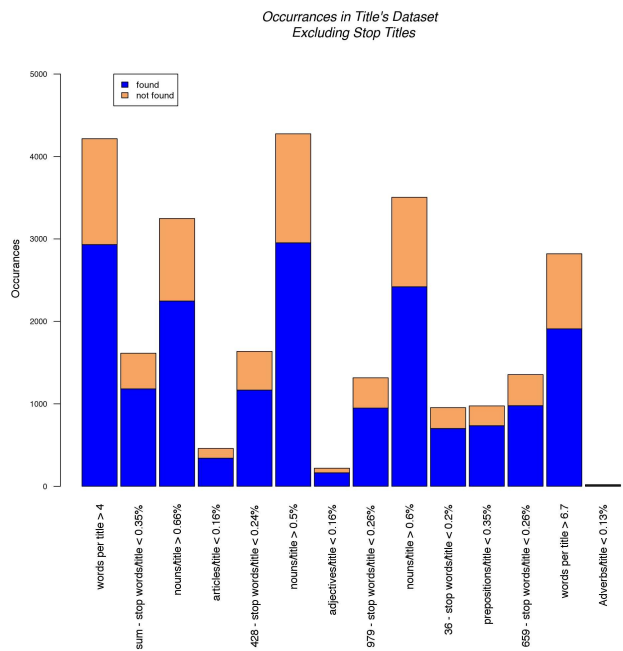

**Fig. 4.** Distribution of Attributes in Titles Over the set of Titles (excludes Stop Titles)

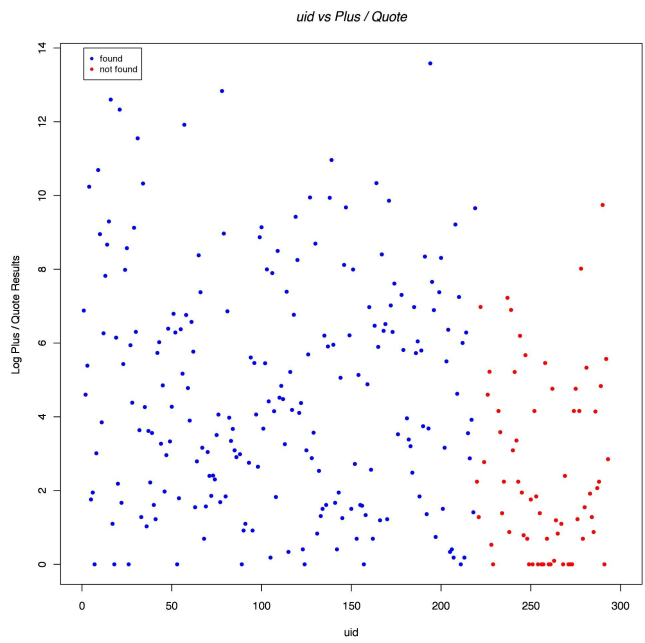

**Fig. 5.** And Searches Divided by Quoted Searches with Respect to Found and Not Found

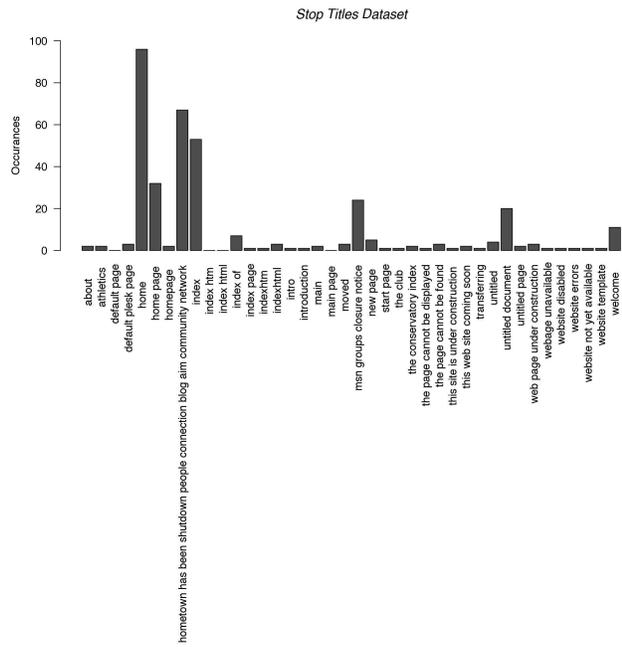

**Fig. 6.** Discovered Stop Titles

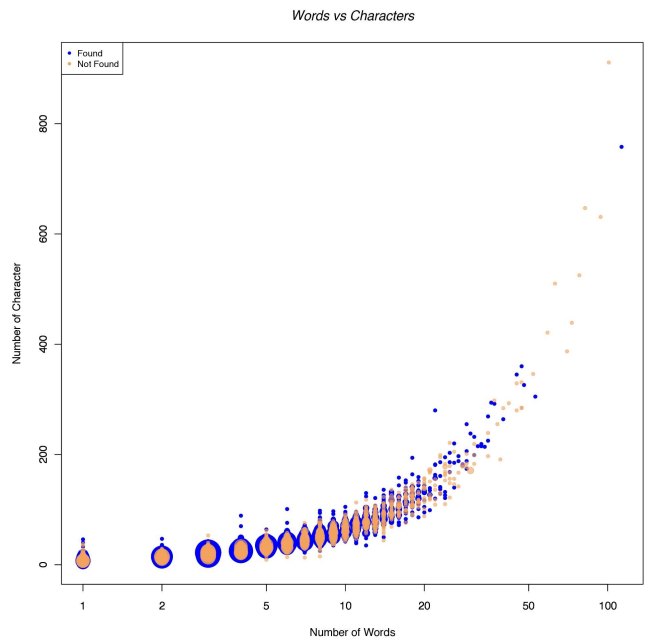

**Fig. 7.** Word versus Characters Characterized by Found and Not Found

## A.2 Charactize Sample Set

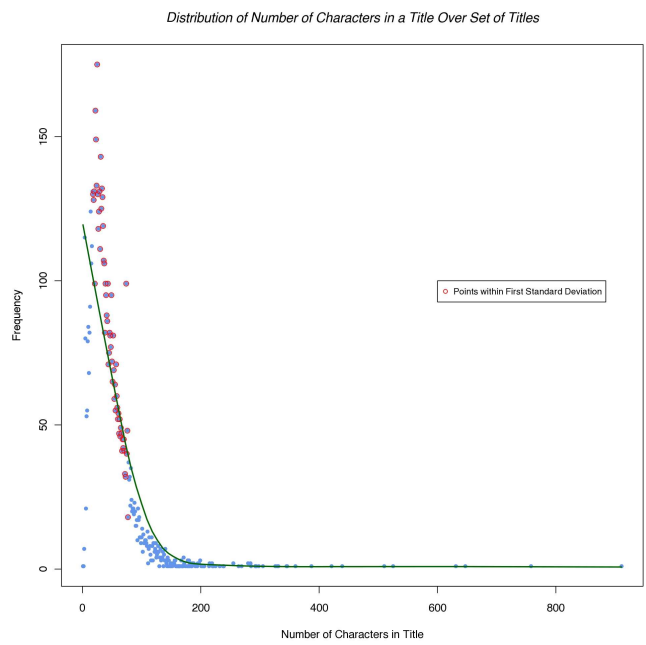

**Fig. 8.** Character Count Frequency Within Set

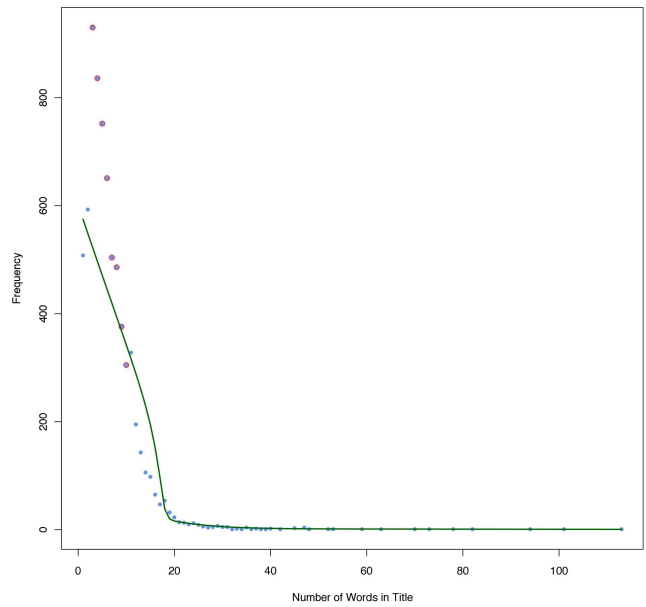

**Fig. 9.** Word Count Frequency Within Set

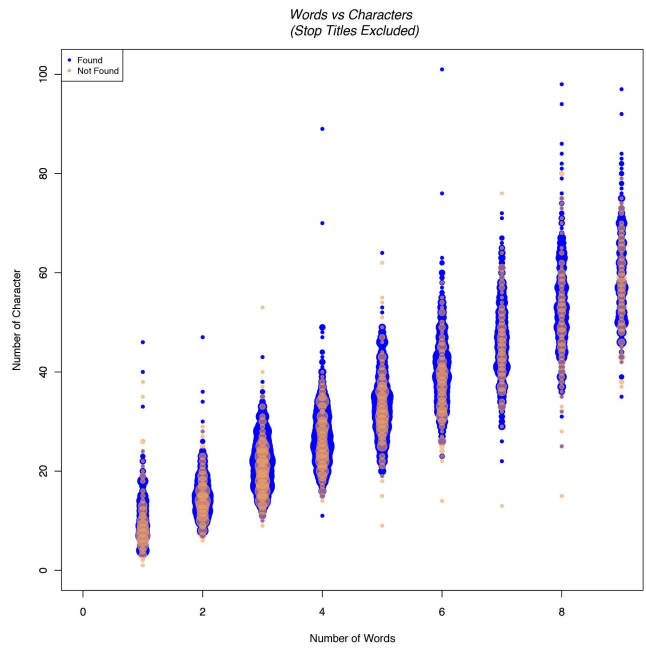

**Fig. 10.** Amount of Characters within Words with Respect to Found and Not Found Searches

Table 12. Table of found and not found data points by word count per title

| count | found | notFound | percentFound | count | found | notFound | percentFound |
|---:|---:|---:|---:|---:|---:|---:|---:|
| 1 | 197 | 127 | 61% | 29 | 4 | 3 | 57% |
| 2 | 363 | 166 | 69% | 30 | 1 | 4 | 20% |
| 3 | 665 | 249 | 73% | 31 | 2 | 3 | 40% |
| 4 | 596 | 210 | 74% | 32 | 1 | 0 | 100% |
| 5 | 546 | 199 | 73% | 33 | 2 | 0 | 100% |
| 6 | 476 | 175 | 73% | 34 | 1 | 0 | 100% |
| 7 | 368 | 136 | 73% | 35 | 2 | 2 | 50% |
| 8 | 357 | 129 | 73% | 36 | 1 | 0 | 100% |
| 9 | 249 | 127 | 66% | 37 | 1 | 1 | 50% |
| 10 | 216 | 89 | 71% | 38 | 0 | 1 | 0% |
| 11 | 175 | 86 | 67% | 39 | 0 | 1 | 0% |
| 12 | 130 | 65 | 67% | 40 | 1 | 1 | 50% |
| 13 | 90 | 53 | 63% | 42 | 0 | 1 | 0% |
| 14 | 75 | 31 | 71% | 45 | 1 | 2 | 33% |
| 15 | 67 | 31 | 68% | 47 | 1 | 3 | 25% |
| 16 | 42 | 23 | 65% | 48 | 1 | 0 | 100% |
| 17 | 29 | 18 | 62% | 52 | 0 | 1 | 0% |
| 18 | 27 | 27 | 50% | 53 | 1 | 0 | 100% |
| 19 | 21 | 11 | 66% | 59 | 0 | 1 | 0% |
| 20 | 10 | 13 | 43% | 63 | 0 | 1 | 0% |
| 21 | 5 | 9 | 36% | 70 | 0 | 1 | 0% |
| 22 | 8 | 5 | 62% | 73 | 0 | 1 | 0% |
| 23 | 6 | 4 | 60% | 78 | 0 | 1 | 0% |
| 24 | 4 | 8 | 33% | 82 | 0 | 1 | 0% |
| 25 | 3 | 6 | 33% | 94 | 0 | 1 | 0% |
| 26 | 3 | 3 | 50% | 101 | 0 | 1 | 0% |
| 27 | 2 | 2 | 50% | 113 | 1 | 0 | 100% |
| 28 | 2 | 3 | 40% | | | | |

Table 13. Characterize Sample Set By Word - First Standard Deviation

| Frequency | Word Count | Percent of Set |
|---|---|---|
| 930 | 3 | 12.9943% |
| 836 | 4 | 11.6809% |
| 752 | 5 | 10.5072% |
| 651 | 6 | 9.09599% |
| 504 | 7 | 7.04206% |
| 486 | 8 | 6.79055% |
| 376 | 9 | 5.2536% |
| 305 | 10 | 4.26156% |

**Table 14.** Characterize Sample Set By Character - First Standard Deviation

| Frequency | Character Count | Percent of Set |
|---|---|---|
| 130 | 18 | 1.8164% |
| 128 | 19 | 1.78846% |
| 131 | 20 | 1.83038% |
| 99 | 21 | 1.38326% |
| 159 | 22 | 2.2216% |
| 149 | 23 | 2.08188% |
| 133 | 24 | 1.85832% |
| 175 | 25 | 2.44516% |
| 130 | 26 | 1.8164% |
| 118 | 27 | 1.64874% |
| 124 | 28 | 1.73257% |
| 131 | 29 | 1.83038% |
| 111 | 30 | 1.55093% |
| 143 | 31 | 1.99804% |
| 125 | 32 | 1.74654% |
| 132 | 33 | 1.84435% |
| 129 | 34 | 1.80243% |
| 119 | 35 | 1.66271% |
| 107 | 36 | 1.49504% |
| 106 | 37 | 1.48107% |
| 82 | 38 | 1.14573% |
| 99 | 39 | 1.38326% |
| 95 | 40 | 1.32737% |
| 88 | 41 | 1.22957% |
| 86 | 42 | 1.20162% |
| 99 | 43 | 1.38326% |
| 71 | 44 | 0.992036% |
| 75 | 45 | 1.04793% |
| 82 | 46 | 1.14573% |
| 81 | 47 | 1.13176% |
| 77 | 48 | 1.07587% |
| 95 | 49 | 1.32737% |
| 72 | 50 | 1.00601% |
| 65 | 51 | 0.908202% |
| 81 | 52 | 1.13176% |
| 69 | 53 | 0.964091% |
| 59 | 54 | 0.824368% |
| 64 | 55 | 0.894229% |
| 55 | 56 | 0.768478% |
| 71 | 57 | 0.992036% |
| 60 | 58 | 0.83834% |
| 56 | 59 | 0.782451% |
| 52 | 60 | 0.726561% |
| 54 | 61 | 0.754506% |

Table 15. Characterize Sample Set By Character - First Standard Deviation

| Frequency | Character Count | Percent of Set |
|---|---|---|
| 47 | 62 | 0.6567% |
| 52 | 63 | 0.726561% |
| 46 | 64 | 0.642727% |
| 49 | 65 | 0.684644% |
| 47 | 66 | 0.6567% |
| 41 | 67 | 0.572866% |
| 45 | 68 | 0.628755% |
| 42 | 69 | 0.586838% |
| 45 | 70 | 0.628755% |
| 41 | 71 | 0.572866% |
| 33 | 72 | 0.461087% |
| 32 | 73 | 0.447115% |
| 99 | 74 | 1.38326% |
| 40 | 75 | 0.558893% |
| 48 | 76 | 0.670672% |
| 18 | 77 | 0.251502% |

## A.3 Confusion Matrix

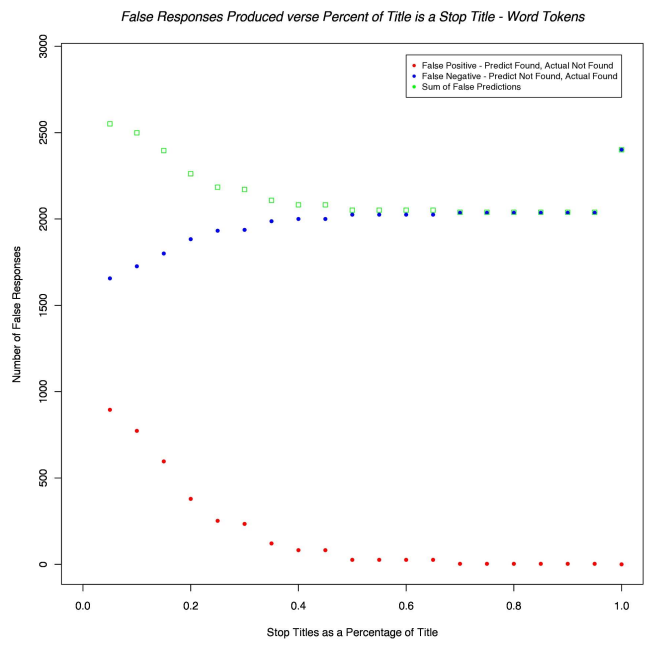

**Fig. 11.** False Responses Produced versus Percent of Title as Stop Title - Word Tokens

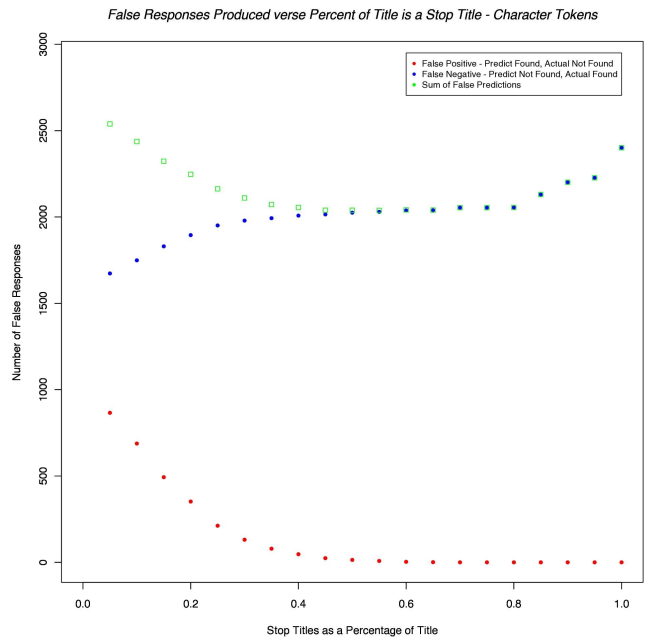

**Fig. 12.** False Responses Produced versus Percent of Title as Stop Title - Character Tokens

**Table 16.** Confusion Matrix Described

|  |  | **Actual** |  |
|---|---|---|---|
|  |  | **found** | **not found** |
| **Predicted** | **found** | True Positive | False Positive |
|  | **not found** | False Negative | True Negative |

**Table 17.** Set of adjective stop words divided by number of words less than 0.16

|  | Actual | | Total Mismatch | Percent Mismatch |
|---|---|---|---|---|
|  | found | not found | 2504 | 35 |
| **Predicted** found | 4592 | 164 | | |
| not found | 2340 | 61 | Match | Match |
|  |  |  | 4653 | 65% |

**Table 18.** Set of article stop words divided by number of words less than 0.16

|  | Actual | | Total Mismatch | Percent Mismatch |
|---|---|---|---|---|
|  | found | not found | 2617 | 37 |
| **Predicted** found | 4415 | 341 | | |
| not found | 2276 | 125 | Match | Match |
|  |  |  | 4540 | 63% |

**Table 19.** 36 stop word divided by number of words less than 0.20

|  | Actual | | Total Mismatch | Percent Mismatch |
|---|---|---|---|---|
|  | found | not found | 2832 | 40 |
| **Predicted** found | 4055 | 701 | | |
| not found | 2131 | 270 | Match | Match |
|  |  |  | 4325 | 60% |

**Table 20.** Set of preposition stop words divided by number of words less than 0.35

|  | Actual | | Total Mismatch | Percent Mismatch |
|---|---|---|---|---|
|  | found | not found | 2883 | 40 |
| **Predicted** found | 4021 | 735 | | |
| not found | 2148 | 253 | Match | Match |
|  | | | 4274 | 60% |

**Table 21.** Set of 659 stop words divided by number of words less than 0.26

|  | Actual | | Total Mismatch | Percent Mismatch |
|---|---|---|---|---|
|  | found | not found | 2968 | 41 |
| **Predicted** found | 3778 | 978 | | |
| not found | 1990 | 411 | Match | Match |
|  | | | 4189 | 59% |

**Table 22.** Set of 979 stop words divided by number of words less than 0.26

|  | Actual | | Total Mismatch | Percent Mismatch |
|---|---|---|---|---|
|  | found | not found | 2949 | 41 |
| **Predicted** found | 3807 | 949 | | |
| not found | 2000 | 401 | Match | Match |
|  | | | 4208 | 59% |

**Table 23.** Set of 428 stop words divided by number of words less than 0.24

|  | Actual | | Total Mismatch | Percent Mismatch |
|---|---|---|---|---|
|  | found | not found | 3049 | 43 |
| **Predicted** found | 3589 | 1167 | | |
| not found | 1882 | 519 | Match | Match |
|  |  |  | 4108 | 57% |

**Table 24.** Number of nouns divided by number of words greater than 0.66

|  | Actual | | Total Mismatch | Percent Mismatch |
|---|---|---|---|---|
|  | found | not found | 3389 | 47 |
| **Predicted** found | 2509 | 2247 | | |
| not found | 1142 | 1259 | Match | Match |
|  |  |  | 3768 | 53% |

**Table 25.** Number of nouns divided by number of words greater than 0.6

|  | Actual | | Total Mismatch | Percent Mismatch |
|---|---|---|---|---|
|  | found | not found | 3411 | 48 |
| **Predicted** found | 2336 | 2420 | | |
| not found | 991 | 1410 | Match | Match |
|  |  |  | 3746 | 52% |

**Table 26.** Number of nouns divided by number of words greater than 0.5

|  | Actual | | Total Mismatch | Percent Mismatch |
|---|---|---|---|---|
|  | found | not found | 3707 | 52 |
| **Predicted** found | 1803 | 2953 | | |
| not found | 754 | 1647 | Match | Match |
|  |  |  | 3450 | 48% |

**Table 27.** Number of words greater than 6.7

|  | Actual | | Total Mismatch | Percent Mismatch |
|---|---|---|---|---|
|  | found | not found | 3823 | 53 |
| **Predicted** found | 1910 | 2846 | | |
| not found | 977 | 1424 | Match | Match |
|  |  |  | 3334 | 47% |

**Table 28.** Number of words greater than 4

|  | Actual | | Total Mismatch | Percent Mismatch |
|---|---|---|---|---|
|  | found | not found | 3975 | 56 |
| **Predicted** found | 1824 | 2932 | | |
| not found | 1043 | 1358 | Match | Match |
|  |  |  | 3182 | 44% |

**Table 29.** Found stop title is URI's title

|  | Actual | | Total Mismatch | Percent Mismatch |
|---|---|---|---|---|
|  | found | not found | 4756 | 66 |
| **Predicted** found | 0 | 4756 |  |  |
| not found | 0 | 2401 | Match | Match |
|  |  |  | 2401 | 34% |

**Table 30.** Found stop title divided by number of words in URI's title less than 3.6E+3

|  | Actual | | Total Mismatch | Percent Mismatch |
|---|---|---|---|---|
|  | found | not found | 4756 | 66 |
| **Predicted** found | 0 | 4756 |  |  |
| not found | 0 | 2401 | Match | Match |
|  |  |  | 2401 | 34% |

**Table 31.** Found stop title divided by number of words in URI's title greater than 0.75

|  | Actual | | Total Mismatch | Percent Mismatch |
|---|---|---|---|---|
|  | found | not found | 2039 | 28 |
| **Predicted** found | 4753 | 3 |  |  |
| not found | 2036 | 365 | Match | Match |
|  |  |  | 5118 | 72% |

**Table 32.** Found stop title divided by number of words in URI's title greater than 0.8

|  | Actual | | Total Mismatch | Percent Mismatch |
|---|---|---|---|---|
|  | found | not found | 2039 | 28 |
| **Predicted** found | 4753 | 3 |  |  |
| not found | 2036 | 365 | Match | Match |
|  |  |  | 5118 | 72% |

**Table 33.** Found stop title divided by number of words in URI's title greater than 0.85

|  | Actual | | Total Mismatch | Percent Mismatch |
|---|---|---|---|---|
|  | found | not found | 2039 | 28 |
| **Predicted** found | 4753 | 3 |  |  |
| not found | 2036 | 365 | Match | Match |
|  |  |  | 5118 | 72% |

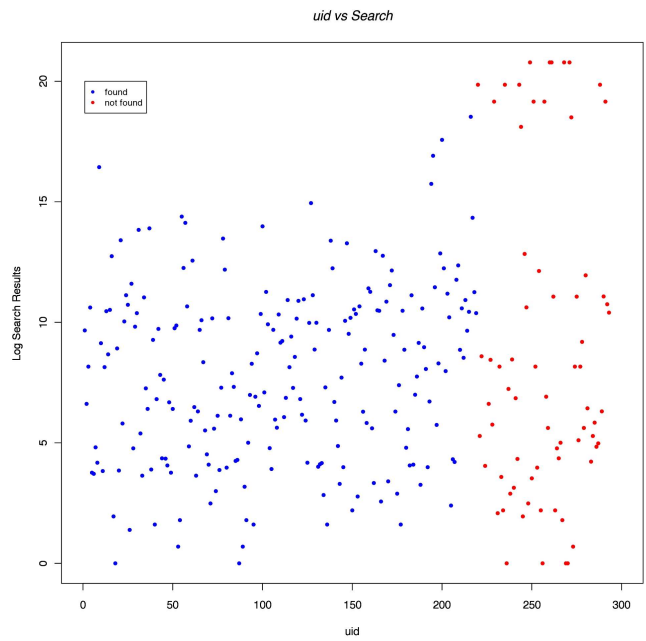

**Fig. 13.** Search Title Results with Respect to Found and Not Found

**Table 34.** Found stop title divided by number of words in URI's title greater than 0.9

|  |  | Actual |  | Total Mismatch | Percent Mismatch |
|---|---|---|---|---|---|
|  |  | found | not found | 2039 | 28 |
| **Predicted** | found | 4753 | 3 |  |  |
|  | not found | 2036 | 365 | Match | Match |
|  |  |  |  | 5118 | 72% |

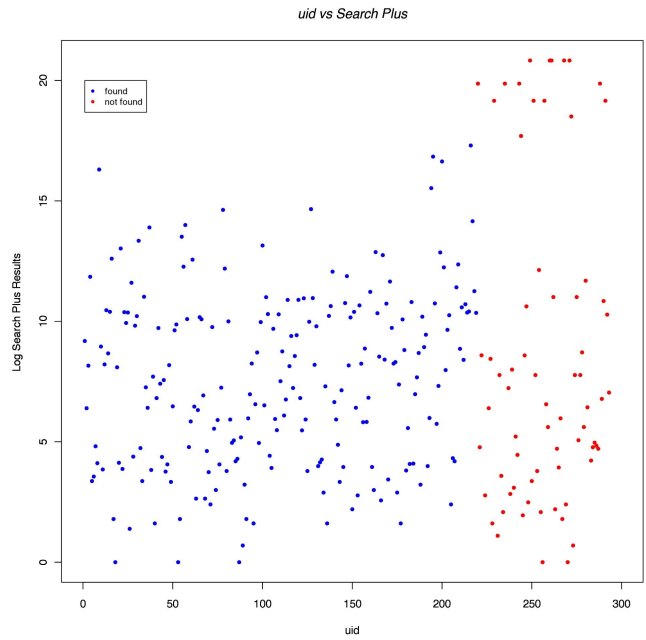

**Fig. 14.** Search Title's "An" Results with Respect to Found and Not Found

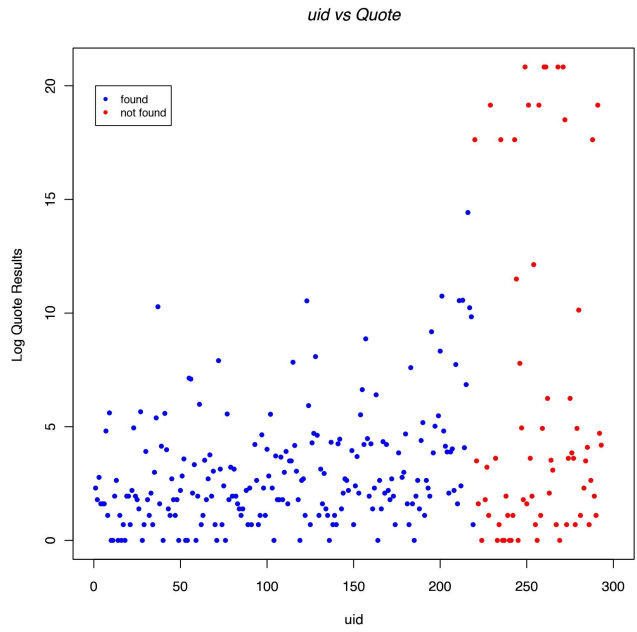

**Fig. 15.** Search Title's "Quote" Results with Respect to Found and Not Found

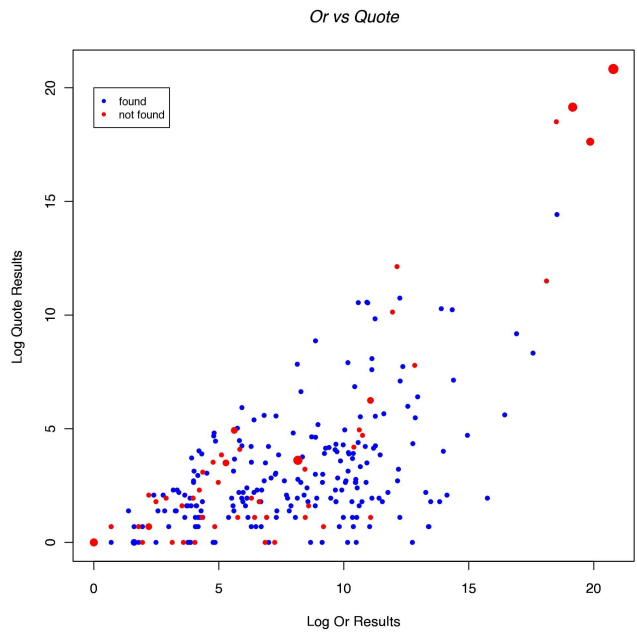

**Fig. 16.** Or Searches versus Quoted Searches with Respect to Found and Not Found

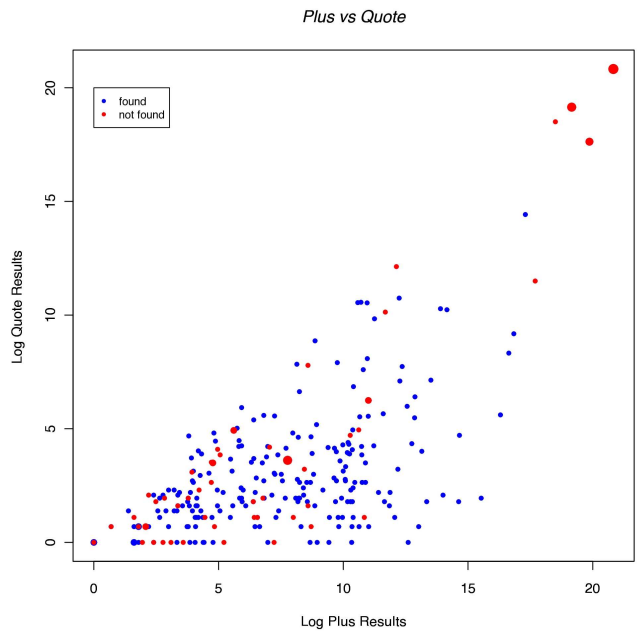

**Fig. 17.** And Searches versus Quoted Searches with Respect to Found and Not Found

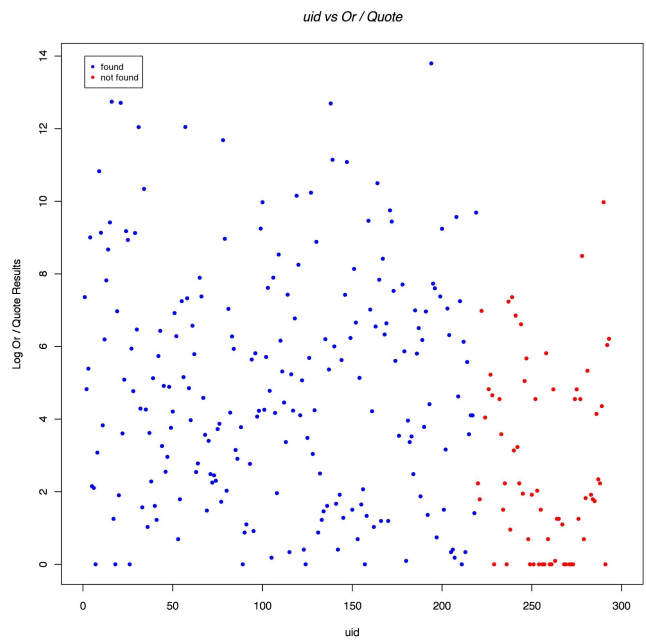

**Fig. 18.** Or Searches Divided by Quoted Searches with Respect to Found and Not Found

**Table 35.** Found stop title divided by number of words in URI's title greater than 0.95

|  |  | **Actual** |  | **Total** Mismatch | **Percent** Mismatch |
|---|---|---|---|---|---|
|  |  | found | not found | 2039 | 28 |
| **Predicted** | found | 4753 | 3 |  |  |
|  | not found | 2036 | 365 | Match | Match |
|  |  |  |  | 5118 | 72% |

**Table 36.** Found stop title divided by number of words in URI's title greater than 0.5

|  |  | **Actual** |  | **Total** Mismatch | **Percent** Mismatch |
|---|---|---|---|---|---|
|  |  | found | not found | 2051 | 29 |
| **Predicted** | found | 4730 | 26 |  |  |
|  | not found | 2025 | 376 | Match | Match |
|  |  |  |  | 5106 | 71% |

**Table 37.** Found stop title divided by number of words in URI's title greater than 0.55

|  |  | **Actual** |  | **Total** Mismatch | **Percent** Mismatch |
|---|---|---|---|---|---|
|  |  | found | not found | 2051 | 29 |
| **Predicted** | found | 4730 | 26 |  |  |
|  | not found | 2025 | 376 | Match | Match |
|  |  |  |  | 5106 | 71% |

**Table 38.** Found stop title divided by number of words in URI's title greater than 0.6

|  | **Actual** |  | Total Mismatch | Percent Mismatch |
|---|---|---|---|---|
|  | found | not found | 2051 | 29 |
| **Predicted** found | 4730 | 26 |  |  |
| not found | 2025 | 376 | Match | Match |
|  |  |  | 5106 | 71% |

**Table 39.** Found stop title divided by number of words in URI's title greater than 0.65

|  | **Actual** |  | Total Mismatch | Percent Mismatch |
|---|---|---|---|---|
|  | found | not found | 2051 | 29 |
| **Predicted** found | 4730 | 26 |  |  |
| not found | 2025 | 376 | Match | Match |
|  |  |  | 5106 | 71% |

**Table 40.** Found stop title divided by number of words in URI's title greater than 0.4

|  |  Actual |  | Total Mismatch | Percent Mismatch |
|---|---|---|---|---|
|  | found | not found | 2082 | 29 |
| **Predicted** found | 4674 | 82 |  |  |
| not found | 2000 | 401 | Match | Match |
|  |  |  | 5075 | 71% |

**Table 41.** Found stop title divided by number of words in URI's title greater than 0.45

|  |  Actual |  | Total Mismatch | Percent Mismatch |
|---|---|---|---|---|
|  | found | not found | 2082 | 29 |
| **Predicted** found | 4674 | 82 |  |  |
| not found | 2000 | 401 | Match | Match |
|  |  |  | 5075 | 71% |

**Table 42.** Found stop title divided by number of words in URI's title greater than 0.35

|  |  Actual |  | Total Mismatch | Percent Mismatch |
|---|---|---|---|---|
|  | found | not found | 2108 | 29 |
| **Predicted** found | 4635 | 121 |  |  |
| not found | 1987 | 414 | Match | Match |
|  |  |  | 5049 | 71% |

**Table 43.** Found stop title divided by number of words in URI's title greater than 0.3

|                    | **Actual** |           | **Total** Mismatch | **Percent** Mismatch |
|--------------------|-----------|-----------|----------|----------|
|                    | found     | not found | 2171     | 30       |
| **Predicted** found | 4522      | 234       |          |          |
| not found          | 1937      | 464       | Match    | Match    |
|                    |           |           | 4986     | 70%      |

**Table 44.** Found stop title divided by number of words in URI's title greater than 0.25

|                    | **Actual** |           | **Total** Mismatch | **Percent** Mismatch |
|--------------------|-----------|-----------|----------|----------|
|                    | found     | not found | 2184     | 31       |
| **Predicted** found | 4504      | 252       |          |          |
| not found          | 1932      | 469       | Match    | Match    |
|                    |           |           | 4973     | 69%      |

**Table 45.** Found stop title divided by number of words in URI's title greater than 0.2

|                    | **Actual** |           | **Total** Mismatch | **Percent** Mismatch |
|--------------------|-----------|-----------|----------|----------|
|                    | found     | not found | 2262     | 32       |
| **Predicted** found | 4377      | 379       |          |          |
| not found          | 1883      | 518       | Match    | Match    |
|                    |           |           | 4895     | 68%      |

**Table 46.** Found stop title divided by number of words in URI's title greater than 0.15

|                    | **Actual** |           | **Total** Mismatch | **Percent** Mismatch |
|--------------------|-----------|-----------|----------|----------|
|                    | found     | not found | 2396     | 33       |
| **Predicted** found | 4160      | 596       |          |          |
| not found          | 1800      | 601       | Match    | Match    |
|                    |           |           | 4761     | 67%      |

**Table 47.** Found stop title divided by number of words in URI's title greater than 1.0

|                    | **Actual** |           | **Total** Mismatch | **Percent** Mismatch |
|--------------------|-----------|-----------|----------|----------|
|                    | found     | not found | 2401     | 34       |
| **Predicted** found | 4756      | 0         |          |          |
| not found          | 2401      | 0         | Match    | Match    |
|                    |           |           | 4756     | 66%      |

**Table 48.** Found stop title divided by number of words in URI's title greater than 0.1

|  | Actual | | Total Mismatch | Percent Mismatch |
|---|---|---|---|---|
|  | found | not found | 2499 | 35 |
| **Predicted** found | 3983 | 773 | | |
| not found | 1726 | 675 | Match | Match |
|  | | | 4658 | 65% |

**Table 49.** Found stop title divided by number of words in URI's title greater than 0.05

|  | Actual | | Total Mismatch | Percent Mismatch |
|---|---|---|---|---|
|  | found | not found | 2551 | 36 |
| **Predicted** found | 3861 | 895 | | |
| not found | 1656 | 745 | Match | Match |
|  | | | 4606 | 64% |

**Table 50.** Found stop title divided by number of characters in URI's title greater than 0.45

|  | **Actual** | | Total Mismatch | Percent Mismatch |
|---|---|---|---|---|
|  | found | not found | 2039 | 28 |
| **Predicted** found | 4732 | 24 | | |
| not found | 2015 | 386 | Match | Match |
|  | | | 5118 | 72% |

**Table 51.** Found stop title divided by number of characters in URI's title greater than 0.5

|  | **Actual** | | Total Mismatch | Percent Mismatch |
|---|---|---|---|---|
|  | found | not found | 2039 | 28 |
| **Predicted** found | 4742 | 14 | | |
| not found | 2025 | 376 | Match | Match |
|  | | | 5118 | 72% |

**Table 52.** Found stop title divided by number of characters in URI's title greater than 0.65

|  | **Actual** | | Total Mismatch | Percent Mismatch |
|---|---|---|---|---|
|  | found | not found | 2040 | 29 |
| **Predicted** found | 4755 | 1 | | |
| not found | 2039 | 362 | Match | Match |
|  | | | 5117 | 71% |

**Table 53.** Found stop title divided by number of characters in URI's title greater than 0.6

|  | Actual | | Total Mismatch | Percent Mismatch |
|---|---|---|---|---|
|  | found | not found | 2041 | 29 |
| **Predicted** found | 4753 | 3 | | |
| not found | 2038 | 363 | Match | Match |
|  | | | 5116 | 71% |

**Table 54.** Found stop title divided by number of characters in URI's title greater than 0.75

|  |  | Actual |  | Total Mismatch | Percent Mismatch |
|---|---|---|---|---|---|
|  |  | found | not found | 2054 | 29 |
| **Predicted** | found | 4756 | 0 |  |  |
|  | not found | 2054 | 347 | Match | Match |
|  |  |  |  | 5103 | 71% |

**Table 55.** Found stop title divided by number of characters in URI's title greater than 0.4

|  |  | Actual |  | Total Mismatch | Percent Mismatch |
|---|---|---|---|---|---|
|  |  | found | not found | 2055 | 29 |
| **Predicted** | found | 4709 | 47 |  |  |
|  | not found | 2008 | 393 | Match | Match |
|  |  |  |  | 5102 | 71% |

**Table 56.** Found stop title divided by number of characters in URI's title greater than 0.8

|  |  | Actual |  | Total Mismatch | Percent Mismatch |
|---|---|---|---|---|---|
|  |  | found | not found | 2055 | 29 |
| **Predicted** | found | 4756 | 0 |  |  |
|  | not found | 2055 | 346 | Match | Match |
|  |  |  |  | 5102 | 71% |

**Table 57.** Found stop title divided by number of characters in URI's title greater than 0.35

|  |  | Actual |  | Total Mismatch | Percent Mismatch |
|---|---|---|---|---|---|
|  |  | found | not found | 2072 | 29 |
| **Predicted** | found | 4677 | 79 |  |  |
|  | not found | 1993 | 408 | Match | Match |
|  |  |  |  | 5085 | 71% |

**Table 58.** Found stop title divided by number of characters in URI's title greater than 0.3

|  |  | Actual |  | Total Mismatch | Percent Mismatch |
|---|---|---|---|---|---|
|  |  | found | not found | 2110 | 29 |
| **Predicted** | found | 4625 | 131 |  |  |
|  | not found | 1979 | 422 | Match | Match |
|  |  |  |  | 5047 | 71% |

**Table 59.** Found stop title divided by number of characters in URI's title greater than 0.85

|  |  | Actual |  | Total Mismatch | Percent Mismatch |
|---|---|---|---|---|---|
|  |  | found | not found | 2130 | 30 |
| **Predicted** | found | 4756 | 0 |  |  |
|  | not found | 2130 | 271 | Match | Match |
|  |  |  |  | 5027 | 70% |

**Table 60.** Found stop title divided by number of characters in URI's title greater than 0.25

|  |  | Actual |  | Total Mismatch | Percent Mismatch |
|---|---|---|---|---|---|
|  |  | found | not found | 2163 | 30 |
| **Predicted** | found | 4544 | 212 |  |  |
|  | not found | 1951 | 450 | Match | Match |
|  |  |  |  | 4994 | 70% |

**Table 61.** Found stop title divided by number of characters in URI's title greater than 0.9

|  |  | Actual |  | Total Mismatch | Percent Mismatch |
|---|---|---|---|---|---|
|  |  | found | not found | 2201 | 31 |
| **Predicted** | found | 4756 | 0 |  |  |
|  | not found | 2201 | 200 | Match | Match |
|  |  |  |  | 4956 | 69% |

**Table 62.** Found stop title divided by number of characters in URI's title greater than 0.95

|  |  | Actual |  | Total Mismatch | Percent Mismatch |
|---|---|---|---|---|---|
|  |  | found | not found | 2227 | 31 |
| **Predicted** | found | 4756 | 0 |  |  |
|  | not found | 2227 | 174 | Match | Match |
|  |  |  |  | 4930 | 69% |

**Table 63.** Found stop title divided by number of characters in URI's title greater than 0.2

|  |  | Actual | | Total Mismatch | Percent Mismatch |
|---|---|---|---|---|---|
|  |  | found | not found | 2247 | 31 |
| **Predicted** | found | 4404 | 352 |  |  |
|  | not found | 1895 | 506 | Match | Match |
|  |  |  |  | 4910 | 69% |

**Table 64.** Found stop title divided by number of characters in URI's title greater than 0.15

|  |  | Actual | | Total Mismatch | Percent Mismatch |
|---|---|---|---|---|---|
|  |  | found | not found | 2323 | 32 |
| **Predicted** | found | 4263 | 493 |  |  |
|  | not found | 1830 | 571 | Match | Match |
|  |  |  |  | 4834 | 68% |

**Table 65.** Found stop title divided by number of characters in URI's title greater than 1.0

|  |  | Actual | | Total Mismatch | Percent Mismatch |
|---|---|---|---|---|---|
|  |  | found | not found | 2401 | 34 |
| **Predicted** | found | 4756 | 0 |  |  |
|  | not found | 2401 | 0 | Match | Match |
|  |  |  |  | 4756 | 66% |

**Table 66.** Found stop title divided by number of characters in URI's title greater than 0.1

|  | **Actual** | | Total Mismatch | Percent Mismatch |
|---|---|---|---|---|
|  | found | not found | 2437 | 34 |
| **Predicted** found | 4068 | 688 | | |
| not found | 1749 | 652 | Match | Match |
|  |  |  | 4720 | 66% |

**Table 67.** Found stop title divided by number of characters in URI's title greater than 0.05

|  | **Actual** | | Total Mismatch | Percent Mismatch |
|---|---|---|---|---|
|  | found | not found | 2539 | 35 |
| **Predicted** found | 3890 | 866 | | |
| not found | 1673 | 728 | Match | Match |
|  |  |  | 4618 | 65% |